\documentclass[11pt,draftcls,onecolumn,journal]{IEEEtran}
\usepackage{amsmath,amsfonts}
\usepackage{algorithmic}
\usepackage{algorithm}
\usepackage{array}
\usepackage[caption=false,font=normalsize,labelfont=sf,textfont=sf]{subfig}
\usepackage{textcomp}
\usepackage{stfloats}
\usepackage{url}
\usepackage{verbatim}
\usepackage{graphicx}
\usepackage{cite}
\usepackage{hyperref}
\usepackage{booktabs}
\usepackage{tabularx}
\usepackage{xcolor}
\usepackage{soul}
\usepackage{acronym}
\usepackage{enumitem}

\acrodef{AI}{artificial intelligence}
\acrodef{AoI}{age of information}
\acrodef{IoT}{Internet of Things}
\acrodef{ML}{machine learning}
\acrodef{FL}{federated learning}
\acrodef{DFL}{decentralized \ac{FL}}
\acrodef{SDG}{sustainable development goal}
\acrodef{RER}{renewable energy resource}
\acrodef{SNN}{spiking neural network}
\acrodef{LIF}{leaky integrate-and-fire}
\acrodef{CMOS}{complementary metal-oxide semiconductor}
\acrodef{EDP}{energy-delay-product}
\acrodef{RF}{radio-frequency}
\acrodef{XAI}{explainable \ac{AI}}
\acrodef{PS}{parameter server}
\acrodef{TEE}{trusted execution environment}
\acrodef{SHAP}{Shapley additive explanations}
\acrodef{LIME}{local interpretable model-agnostic explanations}
\acrodef{LRP}{layer-wise relevance propagation}
\acrodef{IDS}{intrusion detection system}
\acrodef{O-RAN}{open radio access network}
\acrodef{xApp}{near-real-time RAN intelligent controller application}
\acrodef{rApp}{non-real-time RAN intelligent controller application}
\acrodef{HE}{homomorphic encryption}
\acrodef{SMC}{secure multiparty computation}
\acrodef{ANN}{artificial neural network}

\makeatletter
\def\ps@IEEEtitlepagestyle{%
  \def\@oddfoot{\copyrightnotice}%
  \def\@evenfoot{}%
}
\makeatother

\newcommand{\copyrightnotice}{%
  \begin{minipage}{\textwidth}
  \footnotesize This work has been submitted to the IEEE for possible publication. Copyright may be transferred without notice, after which this version may no longer be accessible.
  \end{minipage}
}

\begin{document}
%
\title{Trustworthy, Explainable, and Sustainable Decentralized Intelligence for 6G Networks
\thanks{This work has been funded by the European Commission through the Horizon Europe/JU SNS project ROBUST-6G (grant no. 101139068) and through the Italian Ministry of University and Research under the Italian National Recovery and Resilience Plan (NRRP) of NextGenerationEU, project SERICS/ISP5G+ (CUP D33C22001300002).}
}
%
\author{Giovanni~Perin, 
Michele~Rossi,
Enrique Tom\'as Mart\'inez Beltr\'an, 
Fernando Torres-Vega,
Jos\'e Mar\'ia Jorquera Valero,
Manuel Gil P\'erez,
Eunjeong~Jeong, 
Nikolaos~Pappas, 
Farah Abed Zadeh,
Chamara Sandeepa,
Bartlomiej Siniarski, Madhusanka Liyanage,
Betül Güvenç Paltun, 
Leyli Karaçay,
Ioannis Pitsiorlas, 
and Marios Kountouris 

}
%
\markboth{Submitted to IEEE Signal Processing Magazine, September~2026}%
{G. Perin \MakeLowercase{\textit{et al.}}: Trustworthy, Explainable, and Sustainable Decentralized Intelligence for 6G Networks}


\maketitle

\makeatletter
\renewenvironment{IEEEbiographynophoto}[1]{%
  \vspace{-0.5ex}
  \setlength{\parindent}{1em}%
  \noindent\textbf{#1}\ %
}{%
  \par\vspace{2ex}
}
\makeatother

\vspace{-8ex}


As 6G networks transition from theoretical frameworks into operational realities, \ac{AI} evolves from an add-on optimization tool into a distributed and interconnected structural layer. Unlike previous network generations that mostly relied on centralized cloud analytics platforms, \mbox{\ac{AI}-native} 6G networks operate across a dynamic, multi-domain edge-cloud continuum where data originates from heterogeneous sources including user devices, radio access networks, sensing infrastructures, and vertical applications. Centralizing this massive volume of data creates severe communication overhead, unacceptable latency bottlenecks, single points of failure, and complex cross-domain governance challenges. 

\begin{figure}
    \centering
    \includegraphics[width=0.88\linewidth]{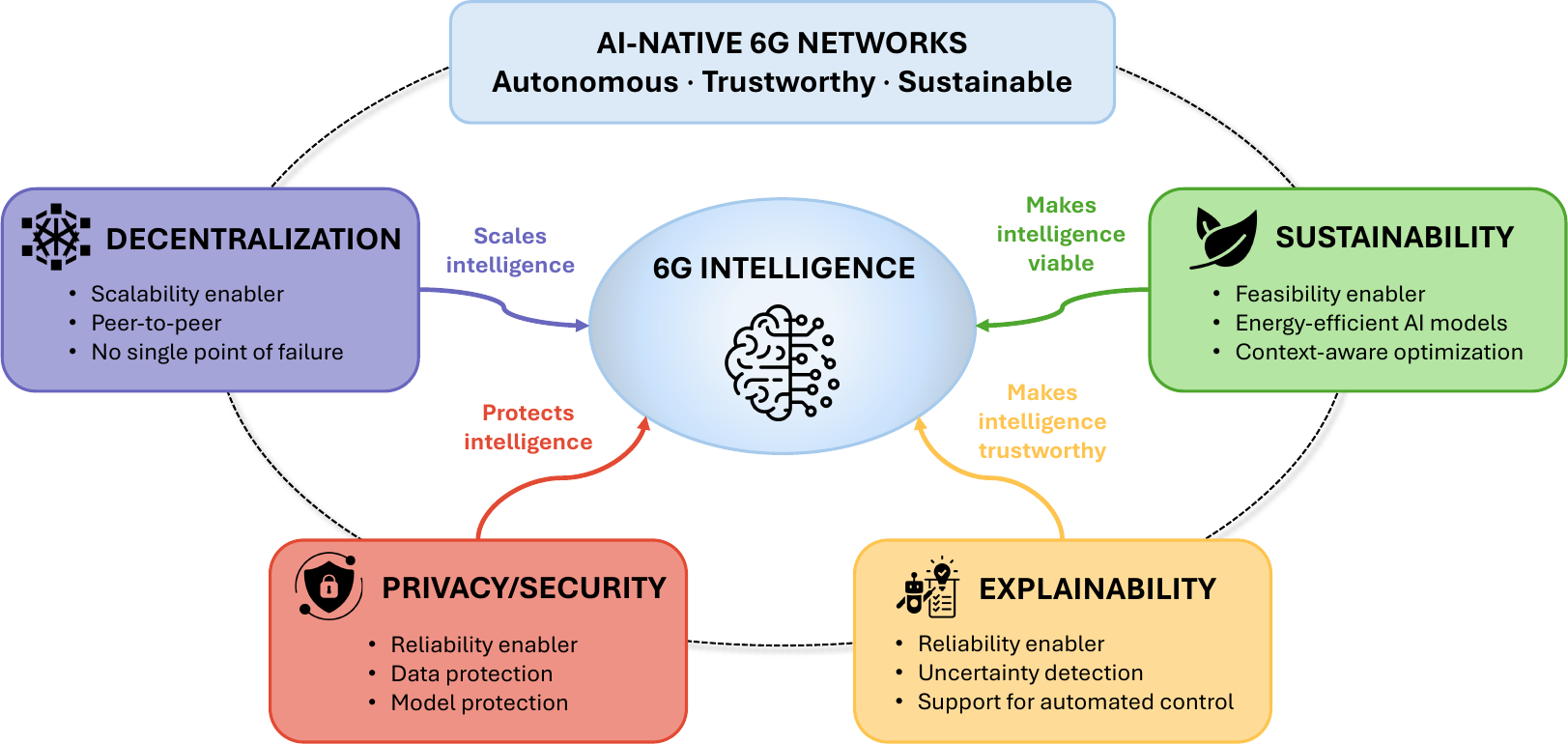}
    \caption{Interdependent design dimensions of intelligence in AI-native 6G networks. Decentralization enables scalable distributed intelligence, privacy/security and explainability provide complementary reliability mechanisms, and sustainability ensures its operational feasibility.}
    \vspace{-0.5cm}
    \label{fig:vision}
\end{figure}

Consequently, {\it decentralization becomes a fundamental architectural requirement} for future 6G network intelligence and zero-touch operations. Security serves as the primary enabler of this decentralized paradigm. Critical security functions, such as real-time threat detection, physical-layer attack mitigation, slice protection, and intrusion detection, require immediate access to local context and telemetry before operational data loses its value. However, moving intelligence to the edge via collaborative paradigms like \ac{FL} and \ac{DFL} introduces complex trade-offs. System security cannot be addressed in isolation; it is deeply intertwined with equally important aspects like {\it trustworthiness}, {\it explainability}, and {\it energy sustainability}. To reliably operate 6G networks within automated control loops, distributed security models must prove resilient against input manipulation, poisoning attacks, and privacy leakage. \Ac{XAI} must evolve from post-hoc explanations intended for human inspection into actionable, machine-readable operational signals that quantify model uncertainty, detect concept drift, and validate peer alerts across administrative domains. Simultaneously, security mechanisms must respect strict energy constraints at edge and \ac{IoT} nodes, but also at base stations and network controllers at the edge boundary, which represent a massive portion of localized energy consumption and hardware footprint. To achieve true edge-to-access sustainability, we advocate combining joint energy- and information-aware distributed training schedulers for \ac{AI} models and implementing such models in dedicated ultra-low-power hardware that is energy efficient by design. For this, we identify neuromorphic platforms involving \acp{SNN} and reservoir computing as the computational means. Figure~\ref{fig:vision} summarizes this vision: decentralization provides the scalability required to distribute intelligence across 6G networks, privacy/security and explainability provide complementary forms of reliability, and sustainability ensures its long-term operational feasibility.

Taking these aspects into account, this paper develops a unified perspective on decentralized intelligence for 6G, arguing that decentralization, trustworthiness, explainability, and sustainability must be designed jointly rather than treated as independent requirements. We start by establishing the architectural transition from centralized machine learning to peer-to-peer decentralized FL, formulating graph-regularized consensus mechanisms that support low-latency threat mitigation across dynamic multi-domain topologies. We then analyze security threats across the AI model lifecycle alongside privacy-preserving cryptographic and hardware-based techniques tailored to distributed environments. Moving beyond traditional explanations, we argue that feature attributions (anomaly scores) and uncertainty metrics can serve as actionable control-plane signals to automatically detect and gauge security threats, driving model self-optimization, policy enforcement, and cross-peer trust verification. Finally, we address energy efficiency as a core design objective, advocating dedicated neuromorphic hardware and semantics-aware scheduling strategies that balance battery states with information freshness in energy-harvesting scenarios.

\section*{Decentralized Intelligence in AI-Native 6G Systems}

We start our discussion by underlining the architectural role of decentralized intelligence in 6G security, discussing why centralized \ac{AI} pipelines are insufficient, how federated and decentralized learning enable collaborative model generation without centralizing raw data, and how distributed intelligence can support closed-loop security management.
\vspace{-0.3cm}
\subsection*{Limitations of Centralized AI for 6G}

Centralized \ac{AI} simplifies the learning pipeline by consolidating data collection, model training, validation, and deployment within a limited number of controlled infrastructure environments. This approach has been effective in many cloud-based analytics systems, but it becomes increasingly restrictive in the context of \ac{AI}-native 6G security.

Its first limitation is {\it scalability}. Future 6G systems will generate large volumes of heterogeneous data from radio, transport, core, edge, application, and sensing domains. Moving this data continuously to a central collector can create excessive communication overhead and energy consumption. The problem becomes particularly relevant for services that require continuous monitoring, rapid model adaptation, or the integration of data from a large number of distributed nodes.

A second limitation is {\it latency}. Many applications requiring real-time decisions must be executed close to the point of observation. Examples in the domain of security include detecting abnormal traffic in a network slice, identifying suspicious radio fingerprints, reacting to jamming or spoofing attempts, or adapting an intrusion detection model at the network edge. In these cases, centralized analysis may introduce delays that are incompatible with the control timescale of the security function. Intelligence that is not timely enough to support mitigation becomes primarily forensic, rather than operational.

A third limitation is {\it governance}. 6G infrastructures will be inherently multi-domain and multi-stakeholder. Operators, vendors, verticals, edge/cloud providers, and service tenants may all contribute to data and computational resources while retaining distinct privacy, regulatory, and business constraints. Under these settings, a centralized data collector or model owner is often unrealistic. Decentralized intelligence enables generating collaborative security models while preserving domain autonomy.

Centralized \ac{AI} also introduces points of concentration in the learning and decision-making pipeline. Data repositories, model-training platforms, aggregation servers, and model registries become critical assets whose failure or compromise may affect large portions of the system. Decentralization reduces such dependency by distributing learning and decision support across multiple nodes, provided that the learning process is designed to operate under heterogeneous, intermittent, and partially trusted participation.
\vspace{-0.3cm}
\subsection*{Distributed Model Training: Federated and Decentralized Learning}

In the path towards decentralization, \ac{FL} provides a natural mechanism for distributed model training in 6G security. Instead of transferring raw data to a centralized repository, each participant trains a local model using its own observations and exchanges model-related information, such as parameters or gradients. Collaborative learning can therefore take place across devices, edge nodes, or administrative domains, while {\it data remain within their original domains}.

For 6G security services, \ac{FL} can support the collaborative training of intrusion detection models, anomaly detectors, traffic classifiers, \ac{RF} fingerprinting models, and attack prediction mechanisms. Local nodes learn from their own operational context, while the overall system benefits from knowledge distributed across multiple domains. This approach is particularly relevant when data sharing is restricted by privacy regulations, commercial policies, or security requirements.

Conventional \ac{FL}, however, commonly relies on a {\it central aggregator}, often referred to as the \ac{PS}. The \ac{PS} coordinates the training process: it receives local updates, combines them into a global model, and redistributes the resulting model to the participants. Although effective in many settings, the implied star topology is not always suitable for 6G. In fact, the \ac{PS} may become a communication bottleneck, a privacy-sensitive entity, an attack target, and {\it a single point of failure}. Moreover, conventional \ac{FL} assumes a level of centralized trust that may not exist in multi-domain deployments.

\begin{figure}[tb]
    \centering
    \includegraphics[width=0.88\linewidth]{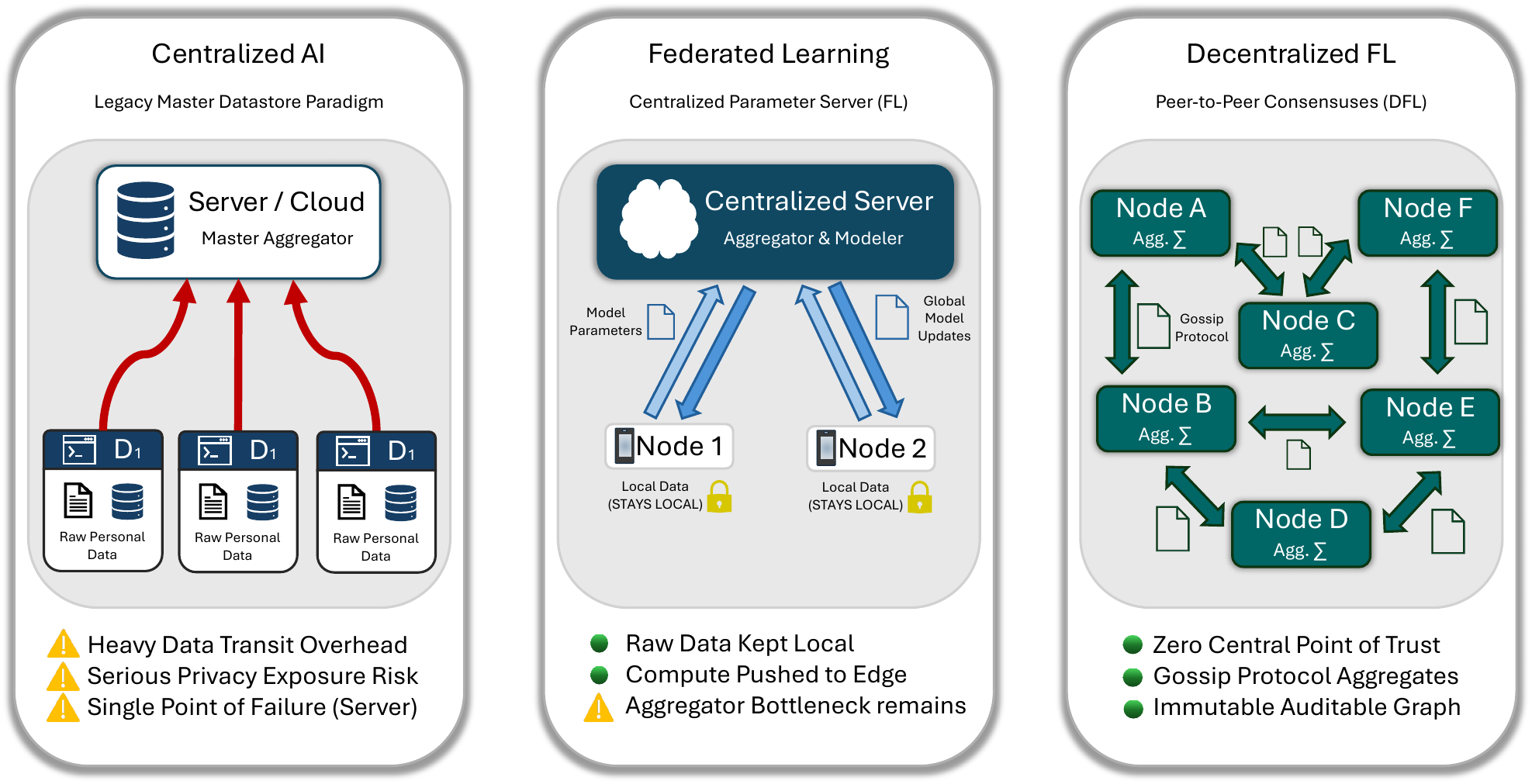}
    \caption{Evolution of AI paradigms from centralized to fully decentralized architectures.
    }
    \label{fig:centralized-fl-dfl}
    \vspace{-0.4cm}
\end{figure}

\Ac{DFL} addresses these limitations by replacing the central aggregator with peer-to-peer collaboration~\cite{beltran23:DFL-survey}. The nodes that take part in the distributed training exchange model updates with selected neighbors according to a potentially time-varying topology. Model training is governed by local optimization, neighborhood-level aggregation, and progressive agreement among the involved peers. The architectural transition from centralized \ac{ML} to \ac{FL} and \ac{DFL} is summarized in Fig.~\ref{fig:centralized-fl-dfl}.

\begin{figure}[tb]
    \centering
    \includegraphics[width=0.88\linewidth]{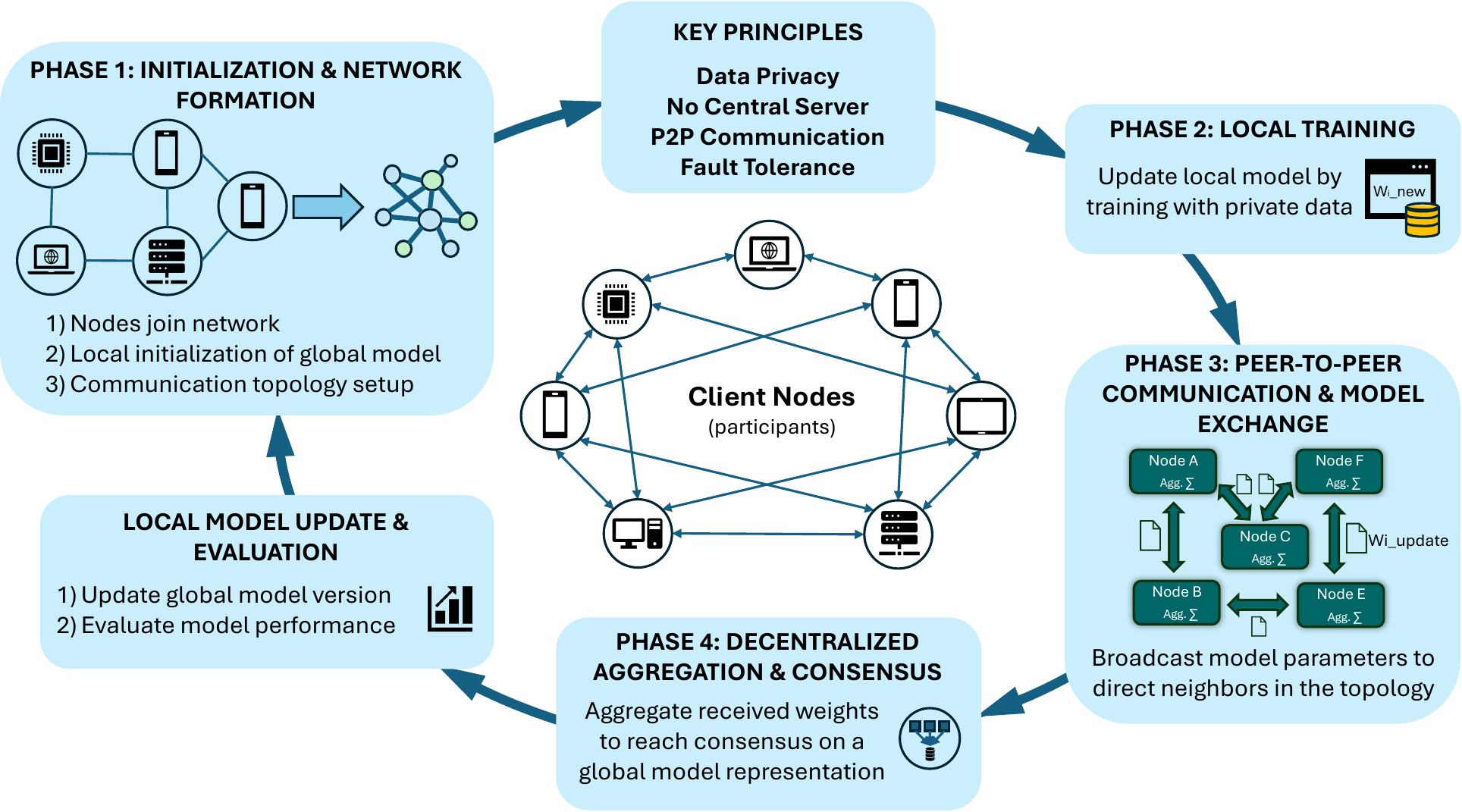}
    \caption{Schematic overview of the DFL closed-loop.
    }
    \label{fig:dfl-closed-loop}
    \vspace{-0.4cm}
\end{figure}

To describe the decentralized learning process mathematically, consider a network of \(K\) learning nodes represented at communication round \(t\) by the weighted graph
\(\mathcal{G}^{t}=(\mathcal{V},\mathcal{E}^{t},\mathbf{A}^{t})\), where
\(\mathcal{V}=\{1,\ldots,K\}\) denotes the set of participating nodes and
\(\mathcal{E}^{t}\) the set of active communication links. An edge
\((k,j)\in\mathcal{E}^{t}\) indicates that nodes \(k\) and \(j\) can exchange model information.
The matrix \(\mathbf{A}^{t}=[a_{kj}^{t}]\) contains the non-negative coupling weights associated with the active links, where \(a_{kj}^{t}\) controls the strength of model agreement between nodes \(k\) and \(j\). Node \(k\) stores a local dataset \(\mathcal{D}_k\), maintains a model \(\mathbf{w}_k^t\) (a set of ``weights'' describing the model being trained), and minimizes the local objective \(F_k(\mathbf{w}_k\mid\mathcal D_k)\). The set \(\mathcal{S}_k^t\) contains the neighboring nodes whose updates are available to node \(k\) during round \(t\). A graph-regularized DFL objective function is
\begin{equation}
    \label{eq:dfl-objective}
    \Phi^{t}(\mathbf{W})
=
\sum_{k=1}^{K} p_k F_k(\mathbf{w}_k\mid\mathcal D_k)
+
\frac{\lambda}{2}
\sum_{(k,j)\in\mathcal{E}^{t}}
a_{kj}^{t}
\left\|
\mathbf{w}_k-\mathbf{w}_j
\right\|_2^2,
\end{equation}
where \(\mathbf{W}=\{\mathbf{w}_1,\ldots,\mathbf{w}_K\}\) denotes the collection of local models. The second term in \eqref{eq:dfl-objective} promotes consensus among neighboring models. The coefficients \(p_k\) weight the local objectives, whereas \(\lambda\)  controls the overall strength of the consensus regularization. The edge-specific coefficients \(a_{kj}^{t}\) determine how strongly the models associated with individual active links are encouraged to agree. Representative local-adaptation and robust neighborhood-aggregation steps consistent with this decentralized learning architecture are
\begin{equation}
\label{eq:dfl-iterations}
\begin{aligned}
\texttt{local adaptation: } & \mathbf{z}_{k}^{t}
=
\mathbf{w}_{k}^{t}
-
\eta_t
\widetilde{\nabla}
F_k\!\left(
\mathbf{w}_{k}^{t}\mid
\mathcal{B}_k^t
\right),
\\
\texttt{neighborhood aggregation: } & \mathbf{w}_{k}^{t+1}
=
(1-\rho_t)\mathbf{z}_{k}^{t}
+
\rho_t
\operatorname{RobAgg}
\left(
\{\mathbf{z}_{j}^{t}\}_{j\in\mathcal{S}_{k}^{t}},
\{\alpha_{kj}^{t}\}_{j\in\mathcal{S}_{k}^{t}}
\right).
\end{aligned}
\end{equation}
The stepsize \(\eta_t\) controls the local adaptation of the weights, where \(\mathcal{B}_k^t\subseteq\mathcal{D}_k\) is the data (mini-batch) available at node \(k\) to compute the local stochastic gradient $\widetilde\nabla F_k$. The local adaptation step can be repeated for multiple local iterations with different mini-batches before setting the value $\mathbf z_k^t$ and using it for the aggregation step.
The coefficient \(\rho_t\) balances local adaptation and neighborhood collaboration, whereas the operator \(\operatorname{RobAgg}(\cdot)\) represents a robust aggregation mechanism
selected according to the deployment and threat model. More generally, the aggregation mechanism in~$\eqref{eq:dfl-iterations}$ provides a natural interface between decentralized learning and the trustworthiness and sustainability requirements discussed throughout this paper. The weights $\alpha_{kj}^t$ and the operator $\operatorname{RobAgg}(\cdot)$ need not depend solely on network connectivity, but may account for the \emph{reliability} of individual peers, \emph{uncertainty or anomaly indicators} associated with their updates, \emph{consistency of their explanations} with locally observed evidence, and their current energy and computational \emph{resources}. Hence, neighborhood collaboration can be made {\it context-aware}: model updates are not necessarily treated equally, but can be weighted, filtered, or discarded based on their estimated trustworthiness, informational value, and cost. The \ac{DFL} closed-loop shown in Fig.~\ref{fig:dfl-closed-loop} allows for heterogeneous participation, intermittent connectivity, and multi-domain operation, while adapting the learning process to the topology and operational constraints of the underlying infrastructure.

This peer-to-peer paradigm is particularly well-suited to 6G, where intelligence is expected to span the cloud, edge, far-edge, and device layers without relying on a single coordination point. Model-sharing relationships may be horizontal, e.g., among edge nodes from different domains, or vertical, e.g., between cloud platforms, edge nodes, and constrained devices. \Ac{DFL} thus enables the learning process to follow the topology and operational constraints of the infrastructure, while reducing dependence on centralized aggregation and preserving the autonomy of participating entities.
\vspace{-0.3cm}
\subsection*{Distributed Intelligence and Zero-Touch Management}

Importantly, the value of decentralized intelligence extends beyond collaborative model training. Distributed learning can {\it become part of the network control fabric}, supporting continuous observation, adaptation, and mitigation. While conventional \ac{ML} pipelines often treat model training, deployment, monitoring, and orchestration as separate stages, \ac{AI}-native 6G systems can benefit from tighter interactions among these functions: local observations update distributed models, model outputs support network decisions, actions modify the network state, and the resulting changes generate new data distributions that may spur further adaptation.

A {\it zero-touch architecture} therefore places decentralized intelligence at the interface between learning and control across the edge–cloud continuum. Edge and far-edge nodes not only execute inference tasks, but also monitor their local environments, exchange model information and learning evidence with peers, and provide operational signals to orchestration functions. Model updates, anomaly scores, confidence indicators, resource constraints, and trust-related metadata can serve as inputs to zero-touch management mechanisms, triggering actions such as attack mitigation, network reconfiguration, slice protection, access-control enforcement, or network resource allocation.

However, before the outputs of distributed entities can be incorporated into automated control loops, distributed models and learning processes must be assessed against several trustworthiness dimensions. Relevant properties include robustness, privacy, fairness, explainability, accountability, and sustainability, as well as the reliability of nodes and model updates during training. These requirements establish the connection between decentralized intelligence and the trustworthiness mechanisms discussed in the following section.

Realizing this vision under realistic 6G conditions remains challenging. Nodes are heterogeneous in computational capacity, data is non-IID, and channel quality varies over time and across nodes. Participation may be intermittent, adversaries may be adaptive, and privacy- and security-preserving mechanisms may introduce additional overhead. Future \ac{FL} and \ac{DFL} systems must therefore jointly address learning accuracy, communication cost, energy consumption, privacy leakage, robustness, and operational transparency.

\section*{Trustworthiness \& Explainability of AI Models}

\subsection*{Robustness to Manipulated Input} 

AI models in 6G may be exposed to security threats that target different stages of the model lifecycle, from data collection and training to deployment and inference~\cite{papernot2018sok}. 

During {\it training}, poisoning attacks can corrupt the data or gradients to create backdoors or degrade model performance. Poisoning attacks are training-time attacks in which an adversary deliberately manipulates training data, labels, gradients, or model updates to influence the learned model~\cite{wang2022threats}. These attacks are especially challenging to detect in large-scale, heterogeneous 6G environments, where data originates from multiple, potentially untrusted sources with varying quality and statistical properties.
At {\it inference} time, adversarial attacks can subtly manipulate inputs through carefully crafted perturbations that remain imperceptible to humans but cause incorrect model predictions. In the context of 6G, such mispredictions may lead to critical failures, such as incorrect resource allocation, degraded quality of service, or even large-scale network disruptions.

Mitigating poisoning and adversarial input attacks in 6G \ac{AI} systems requires a holistic defense strategy that combines data-centric and model-centric mechanisms across the entire learning pipeline. For poisoning attacks, robust data validation and sanitization techniques are key to filtering out anomalous or malicious inputs before they influence the training process. Anomaly detection methods, including statistical outlier detection, explainability-guided detection, and learning-based approaches, can identify inconsistent updates or suspicious patterns during distributed training. In federated settings, reputation-based filtering and trust scoring of participating nodes can help isolate unreliable contributors, while robust aggregation techniques---such as median-based or trimmed-mean methods~\cite{yin2018byzantine}---limit the impact of corrupted updates on the global model. At the model level, adversarial robustness can be enhanced through techniques such as adversarial training~\cite{yang2023adversarial,shafahi2019adversarial}, where models are explicitly trained with perturbed inputs to improve resilience. Additionally, uncertainty estimation and ensemble learning~\cite{zhao2022boosttree} approaches can reduce overconfidence in predictions, enabling the system to flag potentially manipulated or out-of-distribution data.
\vspace{-0.3cm}
\subsection*{Ensuring Data Privacy}

Privacy-preserving AI refers to a set of techniques designed {\it to limit the exposure of sensitive data} and protect confidential information throughout the whole AI lifecycle~\cite{xu2021privacy}. This protection is particularly important in AI-enabled 6G networks, where massive volumes of sensitive data are continuously generated by users, devices, applications, and network operations. In this context, model privacy refers to protecting the AI model itself and {\it the information that may be inferred from it}. Even when raw training data are not directly accessible, an adversary may exploit model outputs, parameters, gradients, updates, or prediction interfaces to infer sensitive information about the training data, determine whether a specific record was used during training, reconstruct private features, or extract and replicate the model. Thus, both training-data privacy and model confidentiality may be required, depending on the deployment scenario.

Unlike centralized \ac{AI} systems, 6G environments are inherently distributed and multi-stakeholder. Data is generated and stored across edge devices, network functions, and service providers, making direct data sharing impractical or undesirable. As a potential solution, privacy-preserving \ac{AI} techniques aim to enable collaborative learning and inference without exposing raw data and ensuring that sensitive information remains protected. However, achieving strong privacy guarantees while maintaining model accuracy, efficiency, and scalability remains a key challenge.
To address privacy risks, a range of complementary techniques is being explored for 6G \ac{AI} systems. For example, federated learning enables distributed model training by keeping data localized on devices, reducing the direct exposure of sensitive information, while secure aggregation protocols prevent leakage through shared model updates. Cryptographic methods such as \ac{HE} and \ac{SMC} allow computations to be performed without exposing raw data to ensure confidentiality even in untrusted environments, although often at the cost of increased computational overhead. \Acp{TEE} provide hardware-based protection for secure model execution and data processing. In addition, differential privacy introduces controlled noise into data or model updates to limit the risk of information leakage about individual data points. Despite these advances, privacy-preserving mechanisms introduce trade-offs in accuracy, latency, use of computational resources, and scalability, requiring careful integration and optimization to meet the high performance requirements of 6G networks.

While \ac{FL} avoids sharing raw data, model updates themselves can still leak sensitive information about the underlying training data. This risk has motivated the development of integrated privacy-aware learning frameworks that combine federated learning with complementary protection mechanisms. At the same time, a key challenge arises from the need to perform model validation, robustness analysis, and attack detection {\it without exposing model updates in plaintext}. For instance, some approaches leverage encryption techniques to encode model updates and compute similarity between them, enabling anomaly detection while preserving confidentiality. These mechanisms are particularly useful to identify adversarial behaviors, such as poisoning attacks, in collaborative learning environments. Referring to \ac{DFL} architectures, privacy-preserving mechanisms can also be incorporated into peer-to-peer learning to protect model updates without relying on a central aggregator. In this case, to increase privacy, participants can encrypt local model updates in a peer-to-peer manner and jointly perform secure aggregation through collaborative cryptographic protocols. Although this improves privacy and eliminates the single point of failure, it requires coordination and techniques such as key sharing to manage cryptographic keys between peers and enable secure aggregations. 

Overall, implementing privacy-preserving \ac{AI} requires striking a careful balance between protecting sensitive information and meeting the security, scalability, energy, and low-latency requirements of 6G networks. 

\subsection*{Explainability, Transparency, and Trust in 6G AI}


As \ac{AI} becomes a core part of 6G, the ability to understand and trust automated \ac{AI} decisions is essential. In 6G networks, \ac{AI} and \ac{ML} models are used across the entire system, from detecting attacks at the network edge to managing threats at the core with little human involvement. When the reasoning behind a model cannot be examined, it is not possible to detect its errors or understand how it could be misled. It is important to distinguish two related notions: \emph{interpretability} is the degree to which a model is inherently understandable, as with a decision tree or a linear model whose logic can be read directly, whereas \emph{explainability} refers to producing human-understandable, usually post-hoc, accounts of why an otherwise opaque model reached a decision. Put simply, an interpretable model explains itself, while an opaque model requires external explanation. \Ac{XAI} provides this external explanation, making the decisions of \ac{AI} and \ac{ML} models understandable and verifiable.

Two main types of \ac{XAI} methods are popular today. The first type is called {\it model-agnostic XAI}, which works with any model, regardless of its internal structure. Examples include \ac{SHAP}, which assigns a score, or {\it attribution}, to each input feature showing how much it contributed to a prediction; the resulting per-feature scores form an \emph{attribution vector}. Another popular model-agnostic method called \ac{LIME} explains individual predictions by testing how small changes to the input affect the output. The second type is {\it model-specific}. For instance, the \ac{LRP} method traces a prediction back through the layers of a neural network to identify which inputs mattered the most.

Among 6G security functions, intrusion detection is where explainability becomes especially relevant. \Acp{IDS} operate continuously at the network edge, processing large volumes of traffic flows, often via \ac{ML} models, with minimal human oversight. Here, the ability to explain why a flow was flagged or cleared serves two purposes: catching model errors early and maintaining trust in automated security operations.
Most XAI research in intrusion detection only focuses on explaining decisions after a model has made them, primarily for human review. While useful, this is not enough for 6G security, where detection must be fast and consistent across multiple operators.
Instead, XAI should be built into the full life cycle of an \ac{IDS}, namely: selecting the most relevant features, reducing model size, detecting when the model starts behaving differently, and deciding when retraining is needed.
Beyond supporting the \ac{IDS} itself, XAI outputs can serve as structured evidence for automated security systems such as \acp{xApp}, \acp{rApp}, and policy engines in \ac{O-RAN} to trigger responses without human input. However, when explanations are used to make security decisions, they also become a target for attacks. Methods such as SHAP and LIME can be manipulated to hide malicious behavior behind misleading explanations. For this reason, the explanation system must be designed to be just as secure and reliable as the detection model it supports.
As 6G moves toward fully automated operation, explainability {\it must go beyond identifying why a decision was made}. It must also express how confident the model is in that decision, so that automated systems can act safely and appropriately.


Accordingly, explanations can serve a much larger purpose. They can be used not only to justify decisions but also to support model optimization, trust assessment, and automated security operations. The following section develops these ideas further, examining how explanations and the corresponding attributions can function as operational signals that support trustworthy and autonomous \ac{IDS} operation in 6G networks.
\vspace{-0.3cm}
\subsection*{Explanations as Actionable, Trustworthy Signals}



As observed above, most \ac{XAI}-for-\ac{IDS} work stops at post-hoc justification, presenting an attribution vector for a human analyst to read. Yet, an attribution vector is itself a compact, machine-readable summary of the model behavior, and can act not only as an explanation for a human, but as an operational signal inside and around the \ac{IDS}.

To support autonomous operation, however, explainability must be paired with a {\it calibrated} assessment of prediction reliability. In fact, an explanation identifies the factors that influenced an IDS decision, but it does not establish whether that decision is sufficiently dependable to justify a subsequent automated network action. This distinction is critical in AI-native 6G environments, where IDS outputs may directly trigger mitigation procedures, traffic isolation, or resource reallocation with limited human intervention. Feature attributions should therefore be accompanied by confidence or uncertainty estimates that characterize the reliability of the underlying prediction. A promising approach is to derive such uncertainty estimates from the latent representations learned by deep anomaly detection models, for example by evaluating latent-space density, similarity to known training samples, or distance from the learned normal-behavior manifold. Samples located in well-represented regions of the normal manifold may support high-confidence benign decisions, whereas samples clearly separated from it may support high-confidence anomaly decisions. By contrast, observations near the decision boundary or in sparsely represented regions should be treated as uncertain and may require additional validation or human escalation. Combined with feature attributions, these indicators would enable xApps, rApps, and orchestration components to determine not only why a decision was made, but also whether it is sufficiently reliable to justify an automated response.

As a signal, these attributions can drive the detector's own lifecycle. Because they rank features by their contribution to a decision, they provide a principled criterion for \emph{feature reduction}. For example, retaining only the features that genuinely drive detection can shrink the model while removing inputs an attacker could cheaply manipulate, thereby reducing the attack surface. The same idea extends inward to \emph{pruning} in \acp{ANN}, where neurons that contribute little to the important attributions are removed to meet the memory, latency, and energy budgets of edge hardware. Attributions also give an early-warning channel for \emph{concept drift}, the situation in which the statistical relationship between the inputs and the quantity being predicted changes over time, so that a model trained on past traffic gradually becomes stale as attack behavior and normal usage evolve. Such drift is usually caught only indirectly: accuracy-based tests react only after misclassifications have accumulated, and tests on the raw input distribution are blind to shifts that leave the marginal feature statistics unchanged while altering the input--output relationship the model has learned. Attributions offer a more direct view. Tracking the \emph{distribution} of attributions over a sliding window of traffic, and flagging when its statistical distance from a stable baseline grows large, can reveal that the features the model relies on---its reasoning---have shifted before accuracy visibly degrades, prompting timely retraining.

Exposed outside the detector, the same signal becomes evidence for automated security control. A binary verdict authorizes only a coarse response, block or allow, whereas a verdict carrying its top contributing features supports a {\it fine-grained} response. In the latter case, the mitigation measure can be targeted at the features driving an alert, access decisions conditioned on whether those features are user-supplied or operator-controlled, and human escalation occurs when the attribution profile matches no known threat or when an attribution provides evidence of out-of-distribution behavior, but its confidence is not sufficiently high to justify an automated decision. Consumed as a compact, schema-conformant record, such explanations let \acp{xApp}, \acp{rApp}, and policy engines in the \ac{O-RAN} architecture act without a human reading them. This matters most when detection is decentralized: as discussed for \ac{DFL}, when models are trained across operators or edge domains, an attribution record becomes a portable unit of trust, letting a peer's alert be independently checked against a recipient's own model and policy before it is acted upon.

Once explanations and attributions drive security actions, {\it the explainer becomes a target in its own right}, and post-hoc attribution methods are not adversarially robust by construction. For example, in {\it explanation manipulation} attacks, crafted inputs yield arbitrary attributions while leaving the model prediction unchanged. In {\it fairwashing}, a model is made to present benign-looking explanations while relying on other features; and {\it backdoor} attacks embed a trigger during training so that the explainer reports an innocent rationale exactly when the model misbehaves. The consequence is that the trustworthiness properties demanded of the detector, i.e., fidelity, stability, adversarial resilience, privacy, and access control, apply just as strongly to the explainer. Robustness, privacy, and faithfulness, long studied in isolation, become inseparable: before any automated policy is allowed to depend on an attribution, that attribution must be made robust to adversarial inputs, controlled in what it discloses, and faithful to the model at once. Together with the confidence estimates noted above, this yields explanations that are not merely readable but safe to act on.


Two hypotheses follow that we believe deserve investigation. First, because a shift in a model’s reasoning manifests as a shift in its attributions, tracking the {\it distribution of attributions} over a window of traffic may enable {\it earlier} and {\it more reliable} detection of concept drift than tests on the raw input signal. This particularly applies to drifts that leave the input statistics unchanged while altering the learned input--output relationship. The second hypothesis concerns detecting attacks on the explainer itself. Because tampering perturbs the attribution distribution in ways benign traffic does not, monitoring that distribution may expose manipulation once a trustworthy baseline and realistic tamper models have been formalized. Together, these claims reframe \ac{XAI} as a measurable, defensible method rather than a presentation layer, which is most needed by a decentralized, trustworthy 6G security stack.


\begin{table*}[t]
\centering
\caption{Three roles of explanations in a trustworthy, decentralized 6G IDS.}
\label{tab:xai-roles}
\renewcommand{\arraystretch}{1.15}
\begin{tabularx}{\textwidth}{
    >{\hsize=.65\hsize\raggedright\arraybackslash}X
    >{\hsize=.95\hsize\raggedright\arraybackslash}X
    >{\hsize=1.40\hsize\raggedright\arraybackslash}X
}
\toprule
\textbf{Role of \ac{XAI}} &
\textbf{Explanation signal} &
\textbf{What it enables} \\
\midrule
Self-optimization &
Feature and neuron attributions &
Compact, drift-aware detectors tailored to edge constraints \\
Control-plane evidence &
Machine-readable attribution record &
Differentiated, automated response across \acp{xApp}, \acp{rApp}, and peers \\
Assurance &
Attribution stability and integrity &
Explanations that are robust, private, and faithful before policy acts on them \\
\bottomrule
\end{tabularx}
\vspace{-0.4cm}
\end{table*}

\section*{Sustainable Intelligence at the Edge}

Trustworthy decentralized intelligence is operationally meaningful only if its learning, inference, and security mechanisms can be sustained within the energy budgets of edge and far-edge devices. Sustainability must therefore be treated as a system-level design constraint rather than as a post-hoc optimization objective.

Traditionally, energy minimization in networks is treated as the result of an optimization problem, once the system architecture is established. Considering modern 6G systems, this is deemed insufficient. Energy efficiency and sustainability are, in coherence with the \acp{SDG}, objectives that must be pursued right from the design phase in modern networks empowered by \ac{AI} and \ac{ML}. For this reason, two main research directions that consider the decentralization of in-network processing are identified:
\begin{itemize}
    \item The design of \ac{AI}/\ac{ML} models that are inherently energy-efficient and their deployment in ultra-low-power dedicated hardware;
    \item The design of networks whose computing facilities are made sustainable through the exploitation of \acp{RER} and a wise management of energy storage systems and computation scheduling.
\end{itemize}
\vspace{-0.3cm}
\subsection*{Natively Energy Efficient \ac{AI} Models}

Recently, numerous approaches have been devised to make deep learning models more energy-efficient. Among the most notable context-agnostic methods, we mention \emph{pruning} and \emph{quantization}.
\begin{description}[font=\normalfont\itshape]
    \item[Pruning.] Iterative deletion of model parameters (and of the corresponding edges) with absolute values lower than a user-defined threshold, followed by retraining.
    \item[Quantization.] Reduction of the number of bits used to store a model parameter after training.
\end{description}
The state-of-the-art shows that, often, (i) deleting redundant neurons can yield improvements in accuracy and generalization to unseen samples, reducing overfitting risks, and (ii) even by using $8$-bit floating-point numbers, the model accuracy is not significantly degraded, while obtaining, on the other hand, significant advantages in terms of memory footprint and computational power. While these methods are, to a certain extent, effective, they belong to the class of {\it post-training optimization approaches} to make memory and computation more efficient. As such, even when carefully engineered, these approaches can only provide limited improvements.

In contrast, we argue for a paradigm shift toward deep learning models that are energy-efficient by design, and not only through post-training operations. For this, we deem {\it neuromorphic computing} especially suitable, specifically (i) {\it reservoir computing} and (ii) {\it \acp{SNN}}.

\subsubsection*{Reservoir computing} It is a neuromorphic computing paradigm particularly suited for temporal signal processing. Unlike conventional deep neural networks, where all parameters are optimized through backpropagation, reservoir computing relies on a fixed recurrent dynamical system, the {\it reservoir}, that projects the input into a high-dimensional temporal representation. Only a shallow output layer is trained, usually via linear regression, by significantly reducing computational complexity and energy consumption. The reservoir naturally retains temporal information through its internal dynamics, making it especially effective for sequential data and time-varying signals. Typical applications include sensor data analysis, traffic forecasting, anomaly detection, and edge intelligence.

Among the most notable approaches are {\it echo state networks} and {\it liquid state machines}. The latter establishes a direct connection with spiking neural networks, since the reservoir itself is composed of recurrently connected spiking neurons.

\subsubsection*{Spiking neural networks} \acp{SNN} are dynamical systems that closely mimic how the human brain works~\cite{fono2026mathematical}. They propagate signals through the emission of \emph{spikes} of current, encoded in the digital domain by binary values, where ``1'' corresponds to the emission of a spike and ``0'' leads to no signal propagation.

\begin{figure}[t]
    \centering
    \begin{minipage}[c][6cm][c]{0.49\linewidth}
        \centering
        \includegraphics[width=\linewidth]{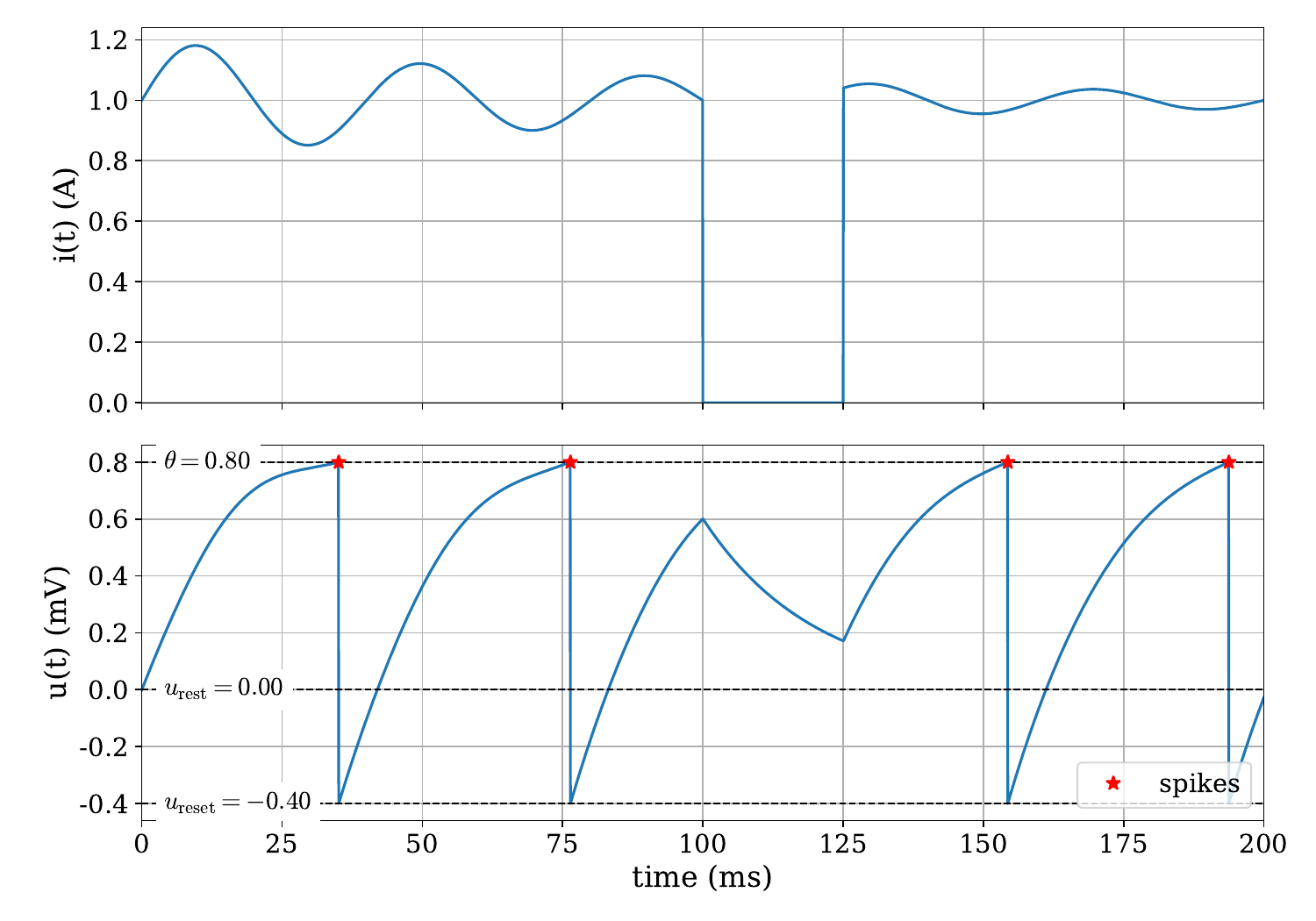}
    \end{minipage}
    \hfill
    \begin{minipage}[c][6cm][c]{0.49\linewidth}
        \centering
        \includegraphics[width=\linewidth]{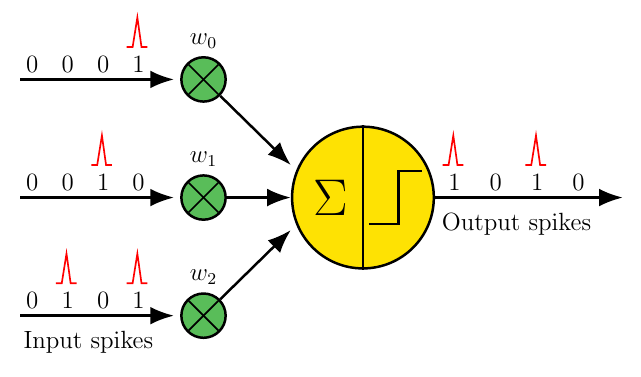}
    \end{minipage}
    \vspace{-0.4cm}
    \caption{On the left: input current $i(t)= 1 + A \sin \left(2\pi\frac{t}{B}\right)\exp\left(-C\,t\right)$ (with $A=0.2$~A, $B=40$~ms, and $C=0.01$~ms$^{-1}$), suppressed for $100 \le t \le 125$ and the corresponding membrane potential $u(t)$ obtained from Eq.~\eqref{eq:snn-membrane-potential} and with the additional mechanisms of the firing and reset thresholds. Here, we used $\tau_m=20$~ms. On the right: schematic representation of a spiking neuron with three inputs (multiplied by the network weights) and the resulting output current.}
    \label{fig:snn-potential}
    \vspace{-0.4cm}
\end{figure}

From a physical perspective, the computationally tractable and most common spiking neuron, the \ac{LIF} neuron, models an $RC$ electric circuit, i.e., the parallel between a resistance $R$ and a capacitor $C$. Neurons in \acp{SNN} are stateful: they dynamically update their \emph{membrane potential} $u(t)$, which represents the neuron's state, through the differential equation
\begin{equation}
    \label{eq:snn-membrane-potential}
    \tau_m \frac{du}{dt} = -\left[u(t)-u_{\rm rest}\right]+R\,i(t),
\end{equation}
where $\tau_m = RC$ is the membrane time constant, affecting the duration of the potential discharging process in time, $u_{\rm rest}$ is the baseline (rest) potential, and $i(t)$ is the input current. This equation tells us that a \ac{LIF} neuron charges its membrane potential when an input current is provided, and exponentially discharges over time (it forgets information). 
In models used in deep learning, two other parameters are defined: a \emph{firing threshold} $\theta$, whose reaching by the membrane potential induces a {\it spike} and a discharge (in the form of a jump) to a {\it reset threshold} $u_{\rm reset}$. A graphic representation of the working principles of \acp{SNN} is shown in Fig.~\ref{fig:snn-potential}: on the left, the membrane potential dynamics as a function of the input current; on the right, the working principle of the spiking neuron in a network, with step function activation. Notably, convolutional, graph, and gated spiking networks can also be implemented.

\subsubsection*{Hardware Implementations} Deploying \acp{SNN} in specifically tailored hardware involves realizing chips using dedicated digital, analog, or mixed-signal circuits, with \ac{CMOS} and emerging memory technologies such as memristive devices providing different realizations of stateful neuronal and synaptic computation. Photonic architectures are also used to implement the paradigm of reservoir computing: photonic reservoir computing exploits the ultra-fast dynamics and low propagation losses of optical systems to enable high-throughput and energy-efficient processing of temporal signals.

Recently, neuromorphic \ac{SNN} hardware has attracted substantial interest in both industry and academia, with prominent platforms including IBM TrueNorth, Intel Loihi, and the academic SpiNNaker architecture. The neuromorphic hardware revolution is also led by startups, such as Innatera, Neuronova, and SynSense. Studies conducted on an Intel Loihi processor show that \acp{SNN} running on dedicated hardware can improve the \ac{EDP} by up to \emph{three orders of magnitude}, making them a promising candidate to break the traditional energy vs.\ latency tradeoff~\cite{rueckauer2022nxtf}.
\vspace{-0.3cm}
\subsection*{Energy- and Context-Aware Distributed Learning}

To conclude our discussion, we now consider embedding energy considerations into the learning phase. To start with, we observe that energy efficiency in \ac{FL} and \ac{DFL} is often framed as a communication problem: how to compress, quantize, or schedule model updates so that less data crosses communication links~\cite{jia25:commst, beltran23:DFL-survey}. On edge devices running deep neural networks, however, this framing overlooks a second, often dominant cost. Local training can consume an amount of energy that is larger than that required to transmit the parameters, so the design objective is not only how to move model updates cheaply, but also how to avoid producing updates that were not worth being computed in the first place.

This second consideration becomes central when devices rely on intermittent, ambient energy, such as solar or \ac{RF} energy harvesting. In these systems, every joule spent on a redundant local training operation is a joule unavailable for a future, more informative one. Scheduling participation on battery availability alone can therefore waste harvested energy on updates that contribute little to the global model. This motivates scheduling policies that are informed of both the energy state of a device and the operational context in which its update will be used.

One line of work addresses the energy dimension through {\it the timing of local computation}. Rather than training as soon as a device has the energy for it, devices can be organized into groups that take turns across a learning cycle, with training deferred until shortly before a group's transmission opportunity. Postponing computation in this way avoids training prematurely on stale global models and then idling until the transmission slot. This regularizes the number of active participants per round, smoothing the otherwise erratic dynamics of energy-harvesting systems. The tradeoff is a controlled amount of staleness that the scheduler must keep bounded.

Energy-awareness alone, however, does not tell the scheduler whether an update is worth computing. A device may be well-charged and well-timed yet {\it carry an update that adds little to the global model}, for instance because its local data distribution is already well-represented or because its model has not drifted significantly since the last round. Assessing this requires looking beyond battery level to the informational content of the prospective update.

A useful lens here is information freshness. The \ac{AoI} and its variants provide a principled method for quantifying how stale a remotely observed process has become and how much value a fresh observation carries, 
and version-based notions of age extend naturally to distributed learning by increasing only when an update would carry significantly new information. Guiding participant selection with this metric biases the system toward contributions that are fresh and substantive; the practical approach is that measuring divergence directly in parameter space is itself costly, so lightweight proxies are needed to keep the check affordable on constrained hardware. This can be done, for example, by comparing compact intermediate representations rather than full parameter vectors~\cite{jeong26:EHFL-VAoI}.

Taken together, these observations point to a broader design approach for sustainable distributed learning at the wireless edge: scheduling decisions should be made {\it jointly in the energy and information domains}. Pure energy-awareness conserves resources but risks spending them on updates of marginal value, whereas pure information-awareness identifies valuable updates but cannot guarantee they are feasible to produce. The two perspectives are complementary, and their benefits are most pronounced precisely where distributed learning is hardest: under severe data heterogeneity and scarce energy. A natural extension carries the same logic from {\it whether} a device should train to {\it how much} it should train, measuring the intensity of each contribution to the expected value of the update. Combined with inherently efficient model architectures, such energy-proportional training would close the loop between model-level and system-level sustainability.

\section*{Concluding Remarks}

This paper analyzed the steps required to realize \ac{AI}-native 6G intelligent functions under strict requirements in terms of security, latency, performance, and energy consumption. This necessitates going beyond isolated optimizations, adopting a unified, multi-dimensional design strategy. Such a strategy entails moving intelligence from centralized clouds to peer-to-peer decentralized architectures to overcome bottlenecks in latency, scalability, and cross-domain data governance. Moreover, decentralization alone is deemed insufficient if the underlying intelligence lacks trust. Thus, appropriate security means must be in place to protect decentralized training, provide algorithmic trustworthiness, and ensure data confidentiality. Building a dependable security infrastructure requires treating trust as a core design requirement rather than a post-deployment adjustment. 

Moreover, to support automated decision-making across zero-touch control loops, explainability tools must evolve beyond human-readable summaries into actionable operational signals that quantify prediction uncertainty, detect concept drift, and validate alerts across peers. 

At the same time, this distributed security stack must operate within strict energy budgets across the entire edge-to-access continuum. To achieve this, we propose a twofold approach: we suggest the adoption of low-power neuromorphic computing architectures and the use of {\it joint} energy- and age-of-information-aware training schedulers to ensure that security mechanisms remain computationally viable without draining excessive resources.


The future 6G security stack cannot be designed as decentralized AI complemented by separate security, explainability, and green-computing mechanisms. These dimensions interact and should jointly determine who participates in learning, whose updates are trusted, when models are adapted, whether their decisions are sufficiently reliable to trigger automated actions, and how much energy should be spent producing those decisions. Their joint design is therefore key to realizing autonomous, trustworthy, and sustainable next-generation networks.

\bibliographystyle{ieeetr}
\bibliography{biblio, IEEEabrv}

\section*{Biographies}

\begin{IEEEbiographynophoto}{Giovanni Perin}(giovanni.perin@unibs.it)
    received his Ph.D. degree in Information Engineering from the University of Padova, Italy. He is an Assistant Professor at the University of Brescia, Italy. His research focuses on distributed learning, optimization, and sustainable AI and networks. He is a Member of IEEE.
\end{IEEEbiographynophoto}

\begin{IEEEbiographynophoto}{Michele Rossi}(michele.rossi@unipd.it)
    received his Ph.D. degree in Information Engineering from the University of Ferrara, Italy. He is a Full Professor at the University of Padova, Italy. His research focuses on sustainable \ac{AI}, wireless sensing, and next-generation networking. He is a Senior Member of IEEE.
\end{IEEEbiographynophoto}

\begin{IEEEbiographynophoto}{Enrique Tom\'as Mart\'inez Beltr\'an}(enriquetomas@um.es)
    received his Ph.D. degree in Computer Science from the University of Murcia, Spain, where he is currently a Postdoctoral Researcher within the CyberDataLab. His research focuses on collaborative learning, large language models, and cybersecurity.
\end{IEEEbiographynophoto}

\begin{IEEEbiographynophoto}{Fernando Torres-Vega}(torresvegaf@um.es)
    is a B.Sc. student in Cybersecurity at the International University of La Rioja, Spain. He is currently a software developer at the University of Murcia, Spain. His research interests include distributed systems, decentralized architectures, and cybersecurity.
\end{IEEEbiographynophoto}

\begin{IEEEbiographynophoto}{Jos\'e Mar\'ia Jorquera Valero}(josemaria.jorquera@um.es) received the M.Sc. and Ph.D. degrees in Computer Science from the University of Murcia. He is currently a Postdoctoral Researcher with the CyberDataLab, University of Murcia. His scientific research interests include trust management, cybersecurity, 5G networks, intent-based management, and continuous authentication.
\end{IEEEbiographynophoto}

\begin{IEEEbiographynophoto}{Manuel Gil Pérez}(mgilperez@um.es)
    received the M.Sc. and Ph.D. degrees in Computer Science from the University of Murcia. He is an Associate Professor with the Department of Information and Communication Engineering, University of Murcia, Spain. His scientific activity focuses mainly on cybersecurity, trust and reputation management, and security operations.
\end{IEEEbiographynophoto}

\begin{IEEEbiographynophoto}{Eunjeong Jeong}(eunjeong.jeong@liu.se)
    received her Ph.D. degree in Communication Systems from EURECOM, France. She is currently a Postdoctoral Researcher at Linköping University, Sweden. Her research focuses on federated learning and energy-efficient scheduling. She is a Member of IEEE.
\end{IEEEbiographynophoto}

\begin{IEEEbiographynophoto}{Nikolaos Pappas}(nikolaos.pappas@liu.se)
    received his Ph.D. degree in Computer Science from the University of Crete, Greece. He is an Associate Professor at Linköping University, Sweden. His research interests include semantic communications, age of information, network-level cooperative networks, and performance analysis and stochastic modeling. He is a Senior Member of IEEE.
\end{IEEEbiographynophoto}

\begin{IEEEbiographynophoto}{Farah Abed Zadeh}(farah.abedzadeh@ucdconnect.ie)
    received her M.Sc. degree in Data and Computational Science from University College Dublin, Ireland, where she is currently a Ph.D. student in Computer Science. Her research focuses on explainable AI in next-generation networks.
\end{IEEEbiographynophoto}

\begin{IEEEbiographynophoto}{Chamara Sandeepa}(abeysinghe.sandeepa@ucd.ie)
is a Postdoctoral Research Fellow at University College Dublin, Ireland. His research interests include AI security, explainable AI, federated learning, and intelligent security solutions for next-generation networks.
\end{IEEEbiographynophoto}

\begin{IEEEbiographynophoto}{Bartlomiej Siniarski}
(bartlomiej.siniarski@ucd.ie) received his Ph.D. degree from University College Dublin, Ireland. He is a Researcher and Co-Director of the Network Softwarization and Security Labs (NetsLab), and is involved in several EU-funded research projects. His research interests include 5G/6G networks, IoT, network security, and network management.
\end{IEEEbiographynophoto}

\begin{IEEEbiographynophoto}{Madhusanka Liyanage}(madhusanka@ucd.ie) is a Professor and Ad Astra Fellow at the School of Computer Science, University College Dublin, Ireland, and Director of the Network Softwarization and Security Labs (NetsLab). His research interests include 5G/6G security, artificial intelligence, explainable AI, federated learning, blockchain, and edge computing. He is a Senior Member of IEEE.
\end{IEEEbiographynophoto}

\begin{IEEEbiographynophoto}{Betül Güvenç Paltun}(betul.guvenc.paltun@ericsson.com) is a Senior Researcher at Ericsson Research, Turkey. Her research focuses on trustworthy AI and intrusion detection systems. 
\end{IEEEbiographynophoto}

\begin{IEEEbiographynophoto}{Leyli Karaçay}(leyli.karacay@ericsson.com)
    received her Ph.D. degree in Computer Science and Engineering from the University of Sabanci, Turkey. She is a Senior Security Researcher at Ericsson Research, Turkey. Her research focuses on distributed learning, network security, and trustworthy AI. 
\end{IEEEbiographynophoto}

\begin{IEEEbiographynophoto}{Ioannis Pitsiorlas}(ioannis.pitsiorlas@eurecom.fr) received his M.Sc. degree in Data Science and Engineering from EURECOM, France, where he is currently a Ph.D. candidate. His research focuses on trustworthy and explainable AI.
\end{IEEEbiographynophoto}

\begin{IEEEbiographynophoto}{Marios Kountouris}(marios.kountouris@eurecom.fr)
    received the Ph.D. degree in Electrical Engineering from T\'el\'ecom Paris, France. He is a Full Professor at EURECOM, France, and a Distinguished Researcher at the University of Granada, Spain. His research focuses on communications theory and machine learning for communications. He is a Fellow of IEEE.
\end{IEEEbiographynophoto}

\end{document}